\documentclass[prl,twocolumn,showpacs]{revtex4}
\usepackage{epsfig}
\usepackage{epstopdf}

\usepackage{times}

\begin{document}

\title{Ab-initio description of hole localization and Zhang-Rice singlets in 
one-dimensional doped cuprates}

\author {Alessio Filippetti and Vincenzo Fiorentini}
\affiliation{CNR-INFM SLACS and Department of Physics, University of Cagliari, I-09042 Monserrato (CA), Italy}

\date{}

\begin{abstract} We present the first ab-initio band-theory-based description of spin-compensated polarons 
(known as Zhang-Rice singlets) in a hole-doped cuprate, specifically one-dimensional Ca$_{2+x}$Y$_{2-x}$Cu$_5$O$_{10}$. 
Zhang-Rice singlets are many-particle configurations relevant to the exotic behavior of hole-doped cuprates, 
stemming from spontaneous charge localization. They appear in our case-study material above a threshold doping,  
successively turning the insulating undoped antiferromagnet into a gap insulator, a singlet-rich metallic paramagnet, 
and finally, a singlet-saturated insulating diamagnet.

\end{abstract}

\maketitle

An outstanding challenge in condensed matter theory is the thorough understanding of the electronic structure of 
hole-doped cuprates. A key ingredient of the physical properties of many such materials is thought to be a 
multiparticle electronic state known as spin-compensated polaron or Zhang-Rice singlet (ZRS in the following), 
first introduced by Zhang and Rice \cite{zhang} in the context of underdoped cuprate superconductors. 
A ZRS is in essence a localized, positively-charged spin-singlet state appearing when (a fractions of) a doping 
hole localizes on one or more O's first-neighbor of the Cu$^{2+}$ site of a  CuO$_2$ unit (embedded in e.g. 
a stripe or layer), coupled anti-ferromagnetically to the polarized native Cu hole. 

There is ample evidence of the presence of ZRS in doped cuprates. Beside their observation \cite{brookes} in 
superconductors and their relevant role therein \cite{anderson04}, 
experiments \cite{hayashi,fong,matsuda, chabot} indicate that the prototypical one-dimensional (1D) cuprate 
Ca$_{2+x}$Y$_{2-x}$Cu$_5$O$_{10}$ (henceforth CaYCuO) must be highly populated by ZRS in a specific doping 
range. In this paper we study CaYCuO by a fully first-principles band theory (the pseudo-self-interaction 
correction approach \cite{fh}, already applied to other correlated cuprates \cite{ff05}) theoretically 
interpreting  the experimental observation of ZRS at low doping, and predicting other ZRS related effects at 
doping levels not experimentally attained so far.

We point out that this is the first instance of a first-principles density-functional-theory-based band 
approach predicting or describing the formation of ZRS, which are usually studied within Hubbard-like 
models \cite{zhang} with simplified electronic structure. While often (perhaps questionably) 
dismissed as inessential in 2D cuprates, a realistic material-specific description is unquestionably necessary 
in the context of 1D cuprates. By and large, first-principles calculations have long been absent from 
this arena due to their difficulties in treating strong correlated cuprates \cite{pickett}. 

1D cuprates are chain-like aggregations of weakly coupled chains of CuO$_2$ units. Compared to other 1D 
systems (e.g. GeCuO$_3$ or Sr$_2$CuO$_3$)  CaYCuO has the advantage  \cite{hayashi} of being hole-dopable 
in a wide range. Magnetization and electrical conductivity measurements \cite{hayashi,fong,matsuda, chabot}  
in the range x/n=[0,0.4] (here x is the hole concentration and n the number of CuO$_2$  units per chain, 
and x/n the hole fraction per Cu) show that CaYCuO is an antiferromagnetic (AF) wide-gap Mott insulator at 
x/n=0 and a Curie-Weiss AF insulator with one-dimensional hopping-conductivity at x/n$<$0.3. At x/n$\sim$0.3 
the AF susceptibility spreads out in a broad maximum, the signature of a magnetic chain populated by a high 
density of spinless sites (ZRS). The conductivity is thermally activated with a gap of 0.08 eV. Thus, perhaps 
surprisingly, the system is effectively insulating through all the experimental doping range. No data are 
reported for concentrations higher than x/n=0.4. Here we deliver a microscopic description of CaYCuO up to 
full doping x/n=1, explaining the mechanism of the observed phase transitions, and uncovering further, not 
yet experimentally accessible doping effects.

\begin{figure}
\epsfxsize=5.5cm
\centerline{\epsffile{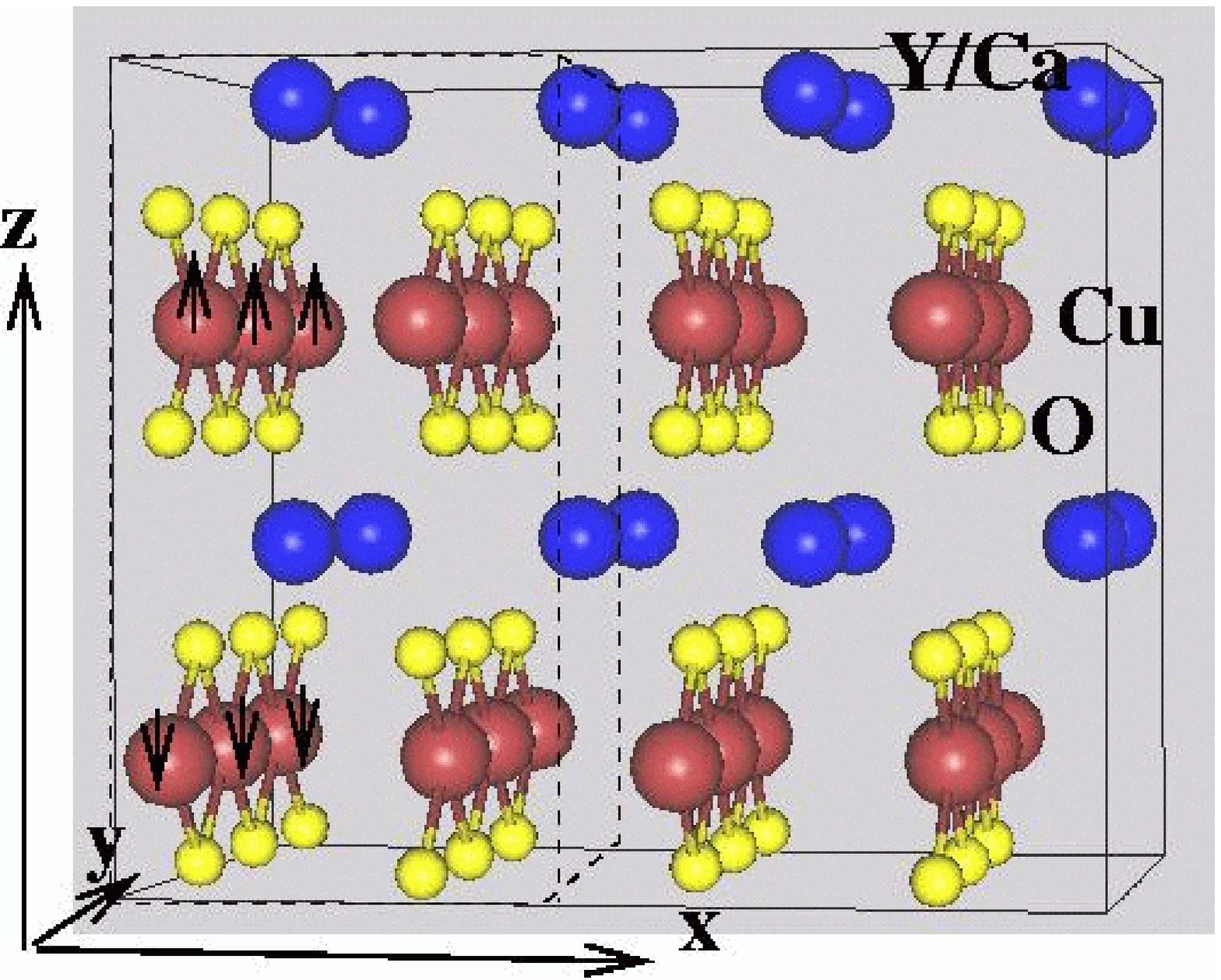}}
\caption{Structure of CaYCuO. FM y-parallel Cu chains couple antiferromagnetically in the z direction. Dashed 
lines delimit the orthorhombic AF unit cell with n=3 (i.e. 3 Cu per unit chain, see text) with lattice 
parameters \cite{fong} $a$=6.172 \AA, $b$/n=2.811 \AA, $c$=10.572 \AA. Arrows show magnetization orientation. 
The ratio of CuO$_2$ to Ca/Y units is n/(n-1).\label{struc}}
\end{figure}

{\it Structure} - CaYCuO (Fig.\ref{struc}) is made of ferromagnetic (FM) CuO$_2$ chains (Cu-Cu separation 
3.4 \AA, Cu-O-Cu angle $\simeq$90$^{\circ}$) running along y and parallel to the (x, z) 
plane \cite{hayashi,fong,matsuda}. Intercalated Ca/Y chains, parallel to the CuO$_2$ ones, act as electron 
reservoir for the CuO$_2$ units. In the synthesized compound Ca$_{2+x}$Y$_{2-x}$Cu$_5$O$_{10}$ the ratio of 
Ca/Y and CuO$_2$ chain steps is 4/5, with n=5.   Contiguous chains are weakly coupled anti-ferromagnetically 
(AF) along z, with T$_{\rm N} $$\sim$28 K at x=0 (16 K at x/n=0.2). Coupling along x is negligible. To gain computational 
flexibility, we treat n as a free parameter (as electronic properties depend on x/n but not on n) and perform 
calculation for n= 2,3,4. We use a plane-wave ultrasoft-pseudopotential method, with 30 Ryd cut-off and 10$\times$10$\times$10 k-point grids; for density of states (DOS) calculations, the linear-tetrahedron interpolation is used on the same mesh. Full details will appear in a forthcoming paper.

\begin{figure}
\epsfxsize=6cm
\centerline{\epsffile{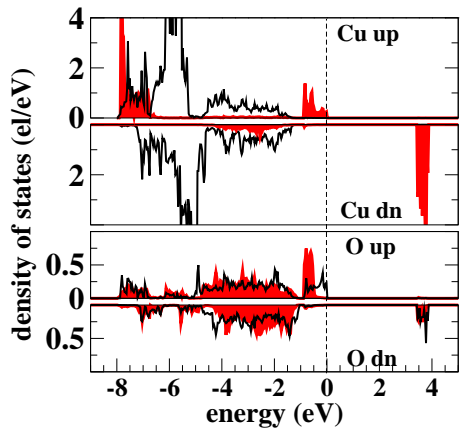}}
\caption{(Color online) Orbital-resolved DOS for undoped CaYCuO. For Cu the contribution of spin-polarized 
d$_{yz}$ orbital is represented by the (red) filled line, while the (black) solid line sums up the other four 
d orbitals. For O, solid and filled lines represent p$_y$ and p$_z$ states, respectively. No Y or Ca DOS  appear 
in this energy range.\label{2}}
\end{figure}

{\it Undoped CaYCuO} -The calculated orbital-resolved DOS in Fig.\ref{2} show that 
undoped (x=0) CaYCuO is a Mott insulator with a  gap of 3.44 eV and Cu magnetic moment of 0.71 $\mu_B$ 
(near-gap bands are shown and discussed in Fig.\ref{4} below). No experimental gap value  is available, but previous studies for low-dimensional cuprates \cite{ff05} suggest our prediction to be sensible. We obtain the observed magnetic ordering as energetically stable, with FM intra-chain exchange-interaction J$_y$=8.5 meV and AF inter-chain J$_z$=--0.2 meV ($\sim$7 meV and $\sim$1 meV  experimentally).

Our calculation attests that CaYCuO is a proper Mott-Hubbard insulator, i.e. valence band top (VBT) and 
conduction band bottom have the same orbital character. All Cu are in the 2+ (d$^9$) state with the highest 
d$_{yz}$ orbital spin-polarized. The flat DOS peak located at VBT and separated from the lower-lying valence 
manifold is due to Cu d$_{yz}$--O (p$_y$, p$_z$) hybrid bands originating from intra-chain FM coupling. 
The magnetization is  about 70\% Cu d$_{yz}$ and 15\% from each adjacent oxygen. The gap opens between 
up- and down-polarized channels of these p-d hybridized bands. It is worth noting that a standard 
density-functional calculation predicts a non-magnetic metal.

\begin{figure}
\epsfxsize=7cm
\centerline{\epsffile{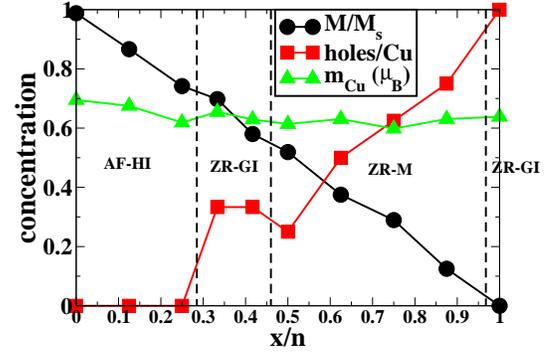}}
\caption{(Color online) Calculated properties as a function of doping. Black circles are relative FM chain 
magnetization per CuO$_2$ unit, green triangles are Cu magnetic moment, red squares are the averaged number 
of ZRS per CuO$_2$ unit. Going from x=0 to x=1 system changes from AF Mott insulator (x=0) to AF hopping 
insulator (HI), ZR gap insulator (GI), ZR metal (M), ZR Mott insulator (at x=1).\label{3}}
\end{figure}

{\it Doped CaYCuO}: Fig.\ref{3} summarizes the calculated properties vs. doping. Fig.\ref{4} reports the 
corresponding band structures.  We find three markedly different regions corresponding to hole concentration 
ranges 0$<$x/n$<$1/4, 1/3$<$x/n$<$1/2, and 1/2$<$x/n$<$1. The x/n=1 case stands out separately, as discussed below.

a) Underdoped region -- At low concentration (0$<$x/n$<$1/4)  injected holes  empty progressively  the 
p-d hybridized VBT peak, without major changes in the band shape: the injected charge disperses along 
the chain, and no ZRS forms. The chain magnetization decreases almost linearly with x, the  AF alignment 
remaining unaffected. As for conduction, Fig.\ref{4} shows that at x/n=1/4 (one hole every four CuO$_2$ 
units) only the very flat (d$_{yz}$--p$_y$, p$_z$) spin-majority bands are cut by E$_F$, and then only 
in the k$_x$, k$_z$-parallel segments M-R and M'-M, orthogonal to the chains.  Therefore the system is 
a low-mobility hopping-conductive insulator. This region is characterized (Fig.\ref{3}) by zero ZRS 
concentration (red squares), a smooth linear decrease of relative chain magnetization (black circles), 
and a weak decrease of Cu magnetic moment (green triangles).

b) ZRS gap insulating region -- As doping nears the threshold x/n=1/3, ZRS populate the system: spin-polarized 
holes localized on the oxygens are coupled with the spin-antialigned holes on the adjacent Cu, so the total 
magnetic moment of the corresponding CuO$_2$ unit is S=0. This occurs because hole-hole Coulomb repulsion 
exceeds the Cu-O charge transfer energy, favoring hole localization. This doping regime exhibits a gapped 
behavior, with transport properties dominated by optical absorption conductivity. Again we note that a 
standard local-density calculation would not obtain hole localization, hence no ZRS.

\begin{figure}
\epsfxsize=9cm
\centerline{\epsffile{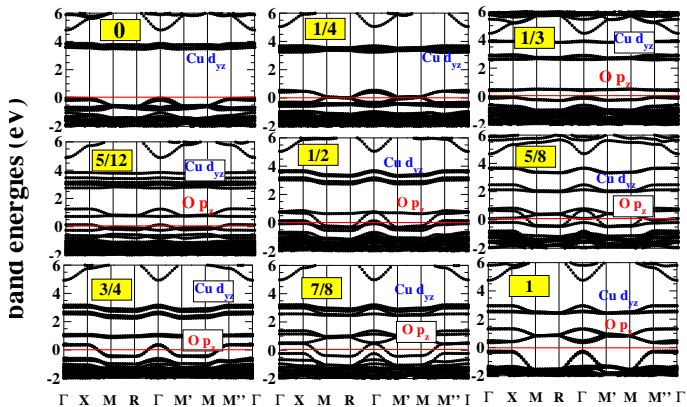}}
\caption{Band energies at various dopings: system is hopping-conductive up to x/n=1/4, gap-insulating 
until x/n=1/2, metallic above x/n=1/2, gap-insulating at both end-points x/n=0 and x/n=1.\label{4}}
\end{figure}

\begin{figure}
\epsfxsize=8cm
\centerline{\epsffile{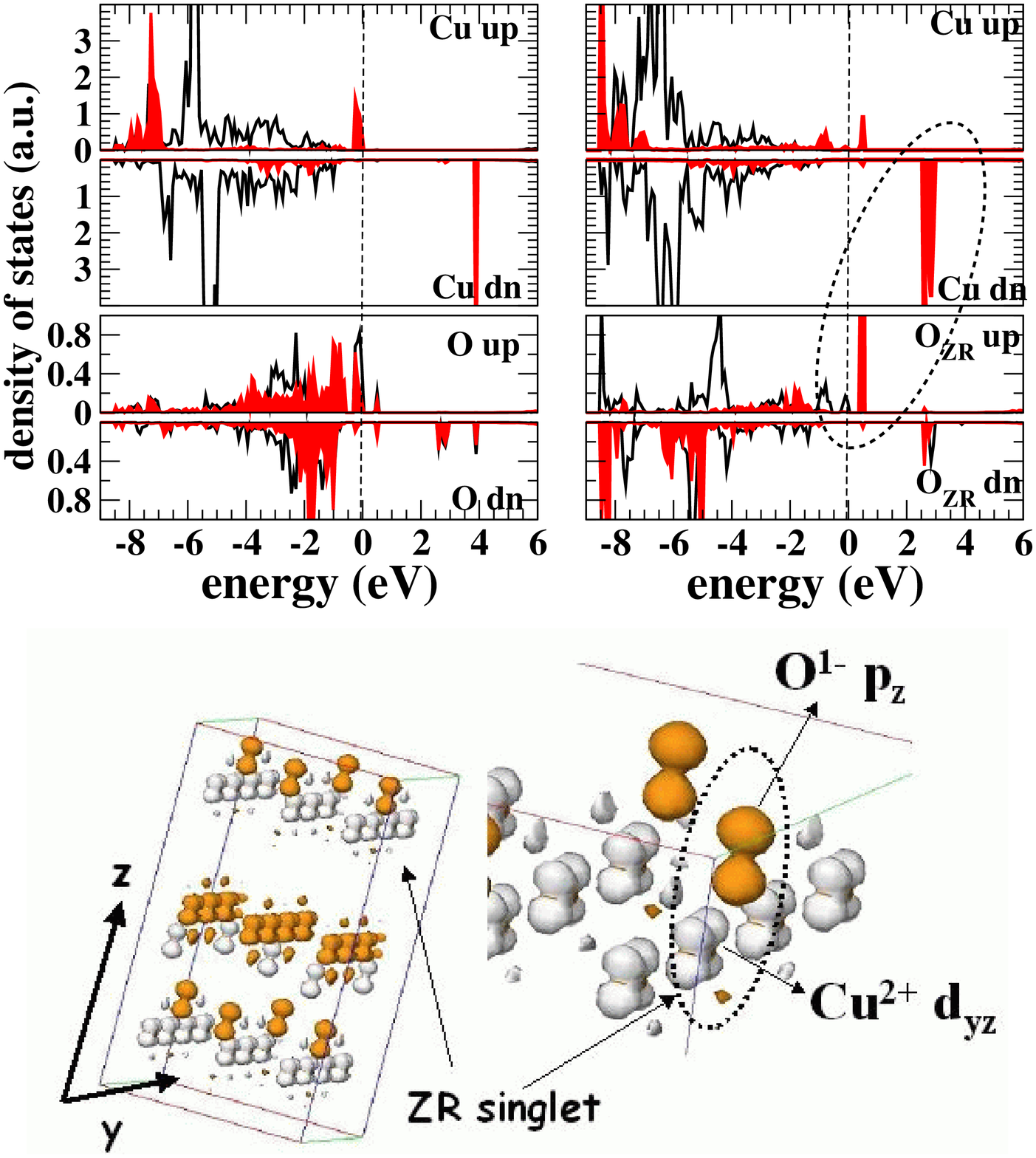}}
\caption{(Color online) Upper panel: orbital resolved DOS of CaYCuO at x/n=1/3 (non-ZRS, left; ZRS, right). 
Holes localize on O p$_z$ states $\sim$0.4 eV above VBT, and couple to adjacent Cu d$_{yz}$ states 
(at $\sim$2.5 eV). The p$_z$ and d$_{yz}$ orbitals (enclosed by ovals) are anti-aligned, i.e. the 
two-particle system is a singlet (ZRS). Lower panel: hole density isosurfaces (h=0.027 bohr$^{-3}$) 
for states within 4 eV above VBT. White and gray (yellow) surfaces are up and down polarized holes. 
In blow-up, the ZRS-forming anti-aligned p$_z$ and d$_{yz}$ orbitals are highlighted by a dashed circle.\label{5}}
\end{figure}

Fig.\ref{5} shows DOS and magnetization isosurfaces for x/n=1/3. At this concentration the holes are 
fully localized, and form one ZRS each three CuO$_2$ steps. Two band features signal the presence of ZRS. 
The first is the flat states of the doping-injected holes, strictly localized on oxygens and with marked 
p$_z $ character, at $\sim$0.4 eV above the VBT. The second signature is the $\sim$1 eV downshift (from 
the undoped value) of part of the Cu d$_{yz}$ spin-minority bands. An antialigned pair of spin-majority 
p$_z$ and spin-minority d$_{yz}$ adjacent holes couple to form one ZRS, whereas the d$_{yz}$ spin-minority 
Cu states not coupled to any localized hole (i.e. not engaged in a ZRS) remain in the undoped-like position 
$\sim$3.5 eV above VBT.

The ZRS is clearly visible in the spin-polarized hole charge isosurfaces of Fig.\ref{5}. Furthermore, 
ZRS formation shows up in the calculated magnetic moments in Table \ref{tab_occ}. Non-ZRS Cu and O atoms 
keep their undoped-state magnetic moment. In the units hosting the localized holes, Cu and O's compensate 
each other quite exactly, forming a ZRS.

\begin{table}
\caption{Orbital contributions to magnetic moments. "Cu, O" refer to non-ZRS units, "Cu$^*$, O$^*$" to ZRS units. 
In the latter, the opposite magnetizations on  d$_{yz}$ and p$_z$ states dominate: the corresponding CuO$_2$ unit 
is demagnetized (S$\simeq$0).}
\label{tab_occ}
\centering
\begin{tabular}{cccccc}
\hline\hline
     M($\mu_B$)  & d$_{yz}$ & p$_x$  &  p$_y$  & p$_z$ & Total\\
\hline
     Cu          &  0.65  &         &        &    &     \\
     O           &        &  0.0    & 0.1    & 0.1&     \\
     Non-ZRS &&&&& 0.67 \\
     \hline
     Cu$^*$         &  0.68  &         &        &      &   \\
     O$^*$          &        &  --0.02  & 0.05   & --0.68 &   \\
     ZRS  &&&&& 0.03 \\
\hline\hline
\end{tabular}
\end{table}

As discussed e.g. in \cite{hayashi,fong, matsuda,chabot}, the low doping-to-ZRS phase transition is associated 
to a change from AF to some disordered paramagnetic (cluster spin-glass or spin-liquid) mixture of ZRS and small 
residual FM chain segments. Indeed, consider in Fig.\ref{3} the difference between CuO$_2$ (black circles) and 
Cu (green triangles) contributions to the average magnetization. The former drops linearly through the whole 
doping range, since holes deplete spin-majority states (either O p or Cu d) at any x/n. Cu magnetization, instead, 
decreases linearly only in the low-doped region, but as soon as ZRS appear, it remains roughly constant up to x/n=1. 
This means that as x increases, ZRS formation demagnetizes more and more CuO$_2$ units, progressively destroying the 
chain FM ordering, while leaving Cu magnetic moments at about their undoped values. (SQUID susceptibility 
data \cite{chabot} also suggest ordered ZRS patterns, but calculations for these structures exceed our present 
computing capabilities.)

c) High-doping ZRS-rich metallic region - As x increases, the number of ZRS increases as long as the hole concentration 
is large enough to allow the "condensation" of more holes on the oxygens. However, as the average hole-hole distance 
is reduced, the charge associated to each ZRS spreads out: at x/n=1/2 (i.e. one ZRS per two CuO$_2$ unit), injected 
holes no longer localize on a single oxygen, but spill out on adjacent CuO$_2$ units. The corresponding states span 
a $\sim$1-eV energy range, much larger than the optical gap of the ZR-GI regime and also larger than the undoped VBT 
DOS peak. As a consequence, the high-doped (x/n$>$1/2) ZRS-rich region is weakly metallic, although large resistivity 
should still be expected due to the small band dispersion. Since metallic ZRS capture a larger hole charge fraction 
than insulating ZRS present at x/n=1/3, at x/n=1/2 some of the chains remain ZRS-free, thus causing the ZRS concentration 
drop visibly in Fig.\ref{3} in correspondence of the metal-insulating transition.

\begin{figure}
\epsfxsize=8.5cm
\centerline{\epsffile{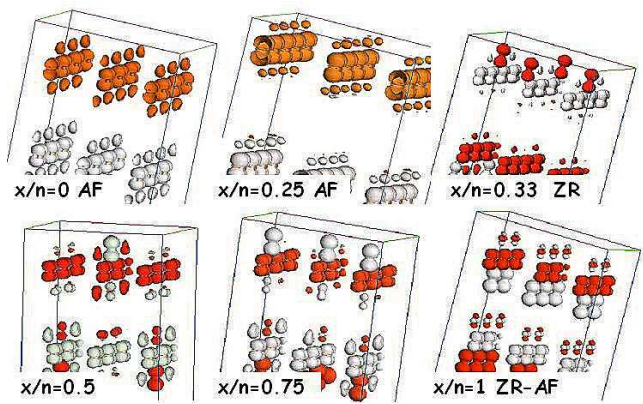}}
\caption{Hole density isosurfaces (same as Fig.5) at various dopings: up to x/n=1/4 the original AF ordering still 
rules; at x/n=1/3 ZRS form, and one each 3 CuO$_2$ units form a singlet; intra-chain FM ordering is lost, and system 
is cluster spin glass or paramagnetic. Above x/n=1/2 more than 50\% of CuO$_2$ has S=0, i.e. system becomes more and 
more non-magnetic.
\label{6}}
\end{figure}

According to the hole isosurface plots in Fig.\ref{6}, in this regime the systems has no recognizable magnetic order, 
and can be seen as a disordered mixture of impure ZRS (i.e. with S not exactly zero: due to metallicity O and Cu 
polarized states do not compensate exactly) and CuO$_2$ units with varying magnetization. (These features will be 
illustrated below in an overview of the evolution of magnetic ordering and band energies vs x/n.)

d) ZRS-saturated diamagnet - The above picture holds for increasing doping, until at x/n=1 (i.e. one hole for each 
CuO$_2$ unit) our calculations predict an exotic ZRS-saturated insulating regime, where each CuO$_2$ hosts one ZRS. 
The system is now a collection of singlet states, hence a non-magnetic (in fact diamagnetic) insulator, despite Cu 
magnetic moments still close to their undoped values. It would be exciting to push the experimental doping limit up 
to x/n=1 and verify the validity of this prediction.

{\it Overview}- In Fig.\ref{6} several frames of hole density isosurfaces are shown as a function of doping. This 
juxtaposition highlights the evolution from low-doping intra-chain FM ordering to ZRS ordering: for x/n$<$1/3 all 
O and Cu holes are parallel-oriented (thus CuO$_2$ magnetization is larger than the Cu magnetic moment) and hole 
isosurfaces can hardly be distinguished from their undoped counterpart at x/n=0. With ZRS formation, holes collapse 
onto O-centered p$_z$-shaped charges, antiparallel (different color in the Figure) to the d$_{yz}$-shaped Cu holes. 
On the other hand, the small hole density residing on non-ZRS oxygens remains spin-parallel (equally colored) to Cu 
holes. The change from ZRS insulating to metallic regime (x/n$\sim$1/2) shows up in the irregular hole-charge 
distributions on the oxygens: now even non-ZRS oxygens can give intra-chain AF contributions to the total chain 
magnetization (because of ZRS broadening to the nearby CuO$_2$ units).

The calculated band structure vs x/n in Fig. \ref{4} confirms again the two regime changes mentioned earlier. 
The transition from low-doping hopping-conductive to ZRS gap-insulating at x/n=1/3 is characterized by the flat 
oxygen bands at $\sim$0.4 eV above the VBT. Thus, in the region 1/3$<$x/n$<$1/2 we expect transport properties to 
be dominated by optical absorption through this energy gap. Above x/n=1/2 the optical gap closes due to ZRS 
spreading out over the chains. At x/n=1 the spectrum is gapped once again: one spin-polarized hole per CuO$_2$ 
is exactly what is needed to empty out the flat p-d bands present at the VBT of the undoped system (see Fig.\ref{2}).

In conclusion, we described ZRS formation and properties in a doped 1D cuprate using a first-principles band theory. 
We interpreted in detail the observed simultaneous change of magnetic and transport properties of CaYCuO upon doping, 
and extended the analysis to dopings not (yet) reachable experimentally. The fact that a first-principles band theory 
proved able to describe ZRS formation in a doped Mott insulator for the first time is an instrumental step for a 
realistic description of a wide class of low-dimensional doped cuprates.

\paragraph*{Acknowledgments}
Work supported in part by MiUR (Italian Ministry of University) through projects PON-Cybersar and PRIN 2005, and by 
Fondazione Banco di Sardegna (Project Correlated oxides).


\begin{thebibliography}{99}
%

\bibitem{zhang}
F. C. Zhang and T. M. Rice, Phys. Rev. B {\bf 37}, 3759 (1988);
H. Eskes and G. A. Sawatzky, Phys. Rev. Lett. {\bf 61}, 1415 (1988).
%

\bibitem{brookes}
N. B. Brookes,  G. Ghiringhelli,  O. Tjernberg, L. H. Tjeng, T. Mizokawa,
T. W. Li, and A. A. Menovsky, Phys. Rev. Lett. {\bf 87}, 237003 (2001).

\bibitem{anderson04}
P. W. Anderson, P. A. Lee, M. Randeria, T. M. Rice, N, Trivedi, and F. C. Zhang, J. Phys.: Condensed Matter {\bf 24}, 
Topical Review R755 (2004).




\bibitem{hayashi}
A. Hayashi, B. Batlogg, and R. J. Cava, Phys. Rev. B {\bf 58}, 2678 (1998), and references therein.

\bibitem{fong}
H. F. Fong, B. Keimer, J. W. Lynn, A. Hayashi, and R. J. Cava, Phys. Rev. B {\bf 59}, 6873 (1999).

\bibitem{matsuda}
M. Matsuda, H. Yamaguchi,T. Ito, C. H. Lee, K. Oka, Y. Mizuno, T. Tohyama, S. Maekawa, and K. Kakurai, 
Phys. Rev. B {\bf 63}, 180403 (2001).


\bibitem{chabot}
M. D. Chabot and J.T. Markert, Phys. Rev. Lett. {\bf 86}, 163 (2000).




Phys. Rev. B 41, 7892 (1990).

\bibitem{fh}
A. Filippetti and N. A. Hill, Phys. Rev. B {\bf 67}, 125109 (2003).  


\bibitem{ff05}
A. Filippetti and V. Fiorentini, Phys. Rev. Lett. {\bf 95}, 086405 (2005); Phys. Rev. Lett. {\bf 98}, 
196403 (2007);  J. Magn. Magnetic Mat. {\bf 310}, 1648 (2007).

\bibitem{pickett}
W. E. Pickett, Rev. Mod. Phys. {\bf 61}, 433 (1989).





\end{thebibliography}
\end{document}